\documentclass[aps,prl,preprint,groupedaddress]{revtex4-1}

\newcommand\TL{\hfil$\displaystyle{##}$}
\newcommand\TR{$\displaystyle{{}##}$\hfil}
\newcommand\TC{\hfil$\displaystyle{##}$\hfil}

\def\seqalign#1#2{\vcenter{\openup1\jot
  \halign{\strut #1\cr #2 \cr}}}

\usepackage{amsmath}
\usepackage{graphicx}
\begin{document}


\title{A New Kind of Scalar Particle}


\author{Christopher G. Tully}
\affiliation{Princeton University}


\date{\today}

\begin{abstract}
The pattern of quantum numbers in the leptons and quarks indicate
that precisely two fixed hypercharge splittings are observed in nature.
The $|\Delta Y|=1$ splitting corresponds to the $SU(2)_L$ scalar and doublet
spacing, indicative of the Yukawa interactions of the Higgs mechanism.  
A $|\Delta Y|=4/3$ splitting is identified between corresponding $SU(3)_C$ scalars
and triplets.  The properties of this splitting are the basis for the
prediction of a new kind of scalar particle that directly couples quarks and
leptons.  Possible experimental methods for detecting a strongly interacting,
charge -1$e$/3 scalar particle are presented for the examples of exotic decay modes
of the top quark and the analysis of $t\bar{t}\tau^+\tau^-$ production 
at hadron colliders.
\end{abstract}


\maketitle

\section{Introduction}
\begin{figure}
\includegraphics[width=\textwidth]{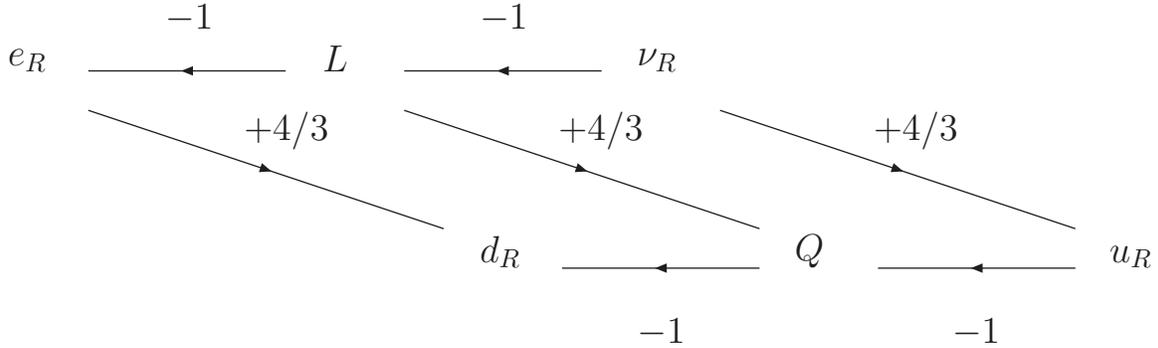}%
\caption{Two hypercharge splittings are observed in the leptons and quarks,
a $|\Delta Y|=1$ splitting along the $SU(2)_L$ singlet to doublet to singlet
transformation and a $|\Delta Y|=4/3$ splitting between corresponding $SU(3)_C$
singlets and triplets.\label{hyperchargesplit}}
\end{figure}

The matter content of the standard model is comprised of three mass generations
of quarks and leptons with quantum number assignments derived from a spontaneously
broken $SU(3)_C \times SU(2)_L \times U(1)_Y$ gauge 
interaction~\cite{standard_model}.
The following list summarizes the ($SU(3)_C$,$SU(2)_L$,$U(1)_Y$) quantum numbers 
of leptons and quarks:
 \begin{equation}
\seqalign{\span\TL\quad\hbox{transforms as}\quad & \span\TR}
{
  u_R & ({\bf 3},{\bf 1},4/3) \ , \cr
  \boldsymbol{Q} = \begin{pmatrix} u_L \\ d_L \end{pmatrix} 
   & ({\bf 3},{\bf 2},1/3) \ , \cr
  d_R & ({\bf 3},{\bf 1},-2/3) \ , \cr
  \boldsymbol{L} = \begin{pmatrix} \nu_L \\ e_L \end{pmatrix}
   & ({\bf 1},{\bf 2},-1) \ , \cr
  \nu_R & ({\bf 1},{\bf 1},0), \cr
  e_R & ({\bf 1},{\bf 1},-2) \ .
 }
 \end{equation}
The assignment of the hypercharge quantum numbers are set according to the known
electric charges of the fermions and are derived from the 
$SU(2)_L \times U(1)_Y \to U(1)_{EM}$ symmetry breaking relationship for the 
electric charge operator:
\begin{equation}
  Q = T^3 + \frac{Y}{2} \ .
 \end{equation}
A particular pattern of hypercharge splitting can be observed between
color-singlet $SU(2)_L$ singlets and doublets in that the magnitude of 
the difference is unity, $|\Delta Y|=1$.  The unity hypercharge splitting
is also present in the difference between color-triplet $SU(2)_L$ 
singlets and doublets.  This splitting is directly related to the
Yukawa interaction terms that generate mass in the fermions via the
Higgs mechanism.
The minimal way to break $SU(2)_L \times U(1)_Y \to U(1)_{EM}$ through the
Higgs mechanism is to have an isospin doublet scalar, the Higgs field 
$\boldsymbol{\Phi}$, with a non-zero vacuum expectation value (vev)~\cite{higgs_mech}:
 \begin{equation}
\seqalign{\span\TC}
{
  \boldsymbol{\Phi} \quad\hbox{transforms as}\quad ({\bf 1},{\bf 2},1) \ , \cr
  \langle\boldsymbol{\Phi}\rangle = \frac{1}{\sqrt{2}} 
   \begin{pmatrix} 0 \\ v \end{pmatrix} \ .
 }
\end{equation}
The hypercharge of the Higgs doublet is set to $Y_\Phi = 1$ so that 
$\langle\boldsymbol{\Phi}\rangle$ is $U(1)_{EM}$-invariant.  Hence, the
value of hypercharge for the Higgs doublet sets the splitting between
chiral components.
Gauge-invariant fermion mass terms appear naturally from spontaneous symmetry breaking
as shown here for the electron,
\begin{equation}
	\Delta{\cal L}_e = - \lambda_e \boldsymbol{\bar{L}} \boldsymbol{\cdot \Phi} e_R + h.c. \ \ .
\end{equation}
If one examines the hypercharge splitting between corresponding $SU(3)_C$ singlets
and triplets, one finds a constant splitting of $|\Delta Y|=4/3$.
The situation is summarized in figure~\ref{hyperchargesplit}.

The observation of this pattern suggests the potential existence of a new
kind of scalar particle, $S$, whose ($SU(3)_C$,$SU(2)_L$,$U(1)_Y$) quantum numbers
are given by
 \begin{equation}
\seqalign{\span\TC}
{
  \boldsymbol{S} \quad\hbox{transforms as}\quad ({\bf 3},{\bf 1},-2/3) \ ,
}
\end{equation}
describing a strongly interacting, charge -1$e$/3 spin-0 boson.
The hypercharge assignment, decribed below, follows from the charge conjugation 
operation in the Dirac current that couples to the $S$ boson, as required by
the selection rules for a scalar interacting with a pair of chiral fermions.

\section{Gauge Supersymmetry}

The pattern of fixed hypercharge splittings 
indicated in figure~\ref{hyperchargesplit} suggests that the leptons and
quarks are part of multiplet with $\Phi$ and $S$ acting as generators of the multiplet.
The Lie group that is a direct product of the Poincar\'e group and the internal
symmetry group of $SU(3)_C \times SU(2)_L \times U(1)_Y$ has a set of 
generators given by $P_\mu$, $M_{\mu\nu}$, $\tau_a/2$, and $\lambda_k/2$
for spacetime translation, rotations, $SU(2)_L$ and $SU(3)_C$ transformations,
respectively.  The addition of the generators $\Phi$ and $S$ into this set
would introduce new symmetries into the Hamiltonian.  The operators $\Phi^\dagger$ 
and $S^\dagger$ replace
$SU(2)_L$ and $SU(3)_C$ singlets with doublets and triplets, respectively.
In the Standard Model, the interactions enter through a gauge-covariant
derivative given by
 \begin{equation}
{\cal D}_\mu = 
\partial_\mu + \frac{i g'}{2} Y B_\mu 
+ i g \frac{\tau_a}{2} W^a_\mu + i g_s \frac{\lambda_k}{2} G^k_\mu \ .
 \end{equation}
We can show that the Hamiltonian is a constant under the replacement
of gauge singlets with doublets or triplets by writing the gauge interaction
explicitly in terms of the anticommutation relation
\begin{equation}
\{\Phi,\Phi^\dagger\} + \{S,S^\dagger\} = 2 H_{int} 
= 2 \gamma^\mu ( - g \frac{\tau_a}{2} W^a_\mu - g_s \frac{\lambda_k}{2} G^k_\mu) \ .
\end{equation}
The introduction of an anticommutation relation to form a graded Lie group
is a known technique of spacetime supersymmetric theories~\cite{susy}.  As the operators
$\Phi$ and $S$ are holding the spin of the leptons and quarks constant, while 
forming a multiplet of gauge singlets and doublets or triplets, the symmetry
operation is a {\it gauge supersymmetry}.  The spin of the gauge bosons and, in general,
the interplay between gauge supersymmetry and spacetime supersymmetry are topics for
further investigation.

In the case of $SU(2)_L$, having the
same symmetry group as spin, the operation of $\Phi$ and $\Phi^\dagger$ transforms
isospin doublets into up-type and down-type singlets and vice versa,
\begin{equation}
\nu_R \xrightarrow{\Phi_1^\dagger} L \xrightarrow{\Phi_2} e_R  \ .
\end{equation} 
In the case of $SU(3)_C$, there are three possible sets of $(\lambda_3,\lambda_8)$
eigenvalues of the fundametal triplet.  We can therefore expect that the operation
of $S$ will take a $Q$ triplet and lower it into three separate $L$ color-singlets.
The triplet multiplicity of observed mass generations of lepton $L$ is therefore 
conjectured as being generated by the three possible gauge sypersymmetric 
operators $S_a$.  For example, one possible cyclic generation of the three
families of lepton and quark isospin doublets could be represented by
\begin{equation}
L_1 \xrightarrow{S_1^\dagger} Q_1 \xrightarrow{S_2} L_2 
\xrightarrow{S_2^\dagger} Q_2 \xrightarrow{S_3} L_3 \xrightarrow{S_3^\dagger} Q_3
\xrightarrow{S_1} L_1 \ .
\end{equation} 

As the gauge supersymmetries described by the operators $\Phi$ and $S$ are not 
respected in nature, the effective couplings of the scalar particles $\Phi$ and $S$
to the leptons and quarks are taken to be of similar magnitude and dependence
as those predicted by the Higgs mechanism.  The mass of the $S$ boson is assumed to
be generated by a coupling between $S$ and the Higgs vacuum expectation value,
giving an observed mass comparable to the electroweak scale, and potentially lighter
than the top quark mass.  The $S$ boson couplings to lepton-quark pairs are assumed to 
have a mass dependence similar to that of a Yukawa interaction term
of a charged Higgs boson in a two-doublet Higgs sector~\cite{HiggsHunters}, as given here:
\begin{equation}
\seqalign{\span\TC}
{
{\cal L}_S = 
({\cal D}_\mu \boldsymbol{S})^\dagger ({\cal D}^\mu \boldsymbol{S})
- \lambda_{\Phi} \boldsymbol{\Phi}^\dagger \boldsymbol{\Phi} \boldsymbol{S}^\dagger \boldsymbol{S}
- \lambda_S ( \boldsymbol{S}^\dagger \boldsymbol{S} ) ^2 \cr
\qquad \qquad \qquad
      - \lambda_L^{ij} \bar{L}_c^{i,k} \epsilon_{kl} Q^{j,l}_a S^*_a
	- \lambda^{R,ij} \left(\bar{\ell}_R\right)_c^i {u}^{j}_{R,a} S^*_a 
      - \lambda_R^{ij} \left(\bar{\nu}_R\right)_c^i {d}^{j}_{R,a} S^*_a + h.c.
}
\end{equation}
where $\lambda_L^{ij}$, $\lambda^{R,ij}$ and $\lambda_R^{ij}$ are general 
complex-valued matrices with $(ij)$-indices over the three mass generations
\begin{equation}
\seqalign{\span\TC}
{
	\boldsymbol{Q}^i = \begin{pmatrix} u_L^i \\ d_L^i \end{pmatrix} 
	= \left( \begin{pmatrix} u_L \\ d_L \end{pmatrix} ,
	\begin{pmatrix} c_L \\ s_L \end{pmatrix} ,
	\begin{pmatrix} t_L \\ b_L \end{pmatrix} \right), \ 
      \boldsymbol{L}^i = \begin{pmatrix} \nu_L^i \\ \ell_L^i \end{pmatrix}
      = \left( \begin{pmatrix} \nu_{e,L} \\ e_L \end{pmatrix} ,
      \begin{pmatrix} \nu_{\mu,L} \\ \mu_L \end{pmatrix} ,
      \begin{pmatrix} \nu_{\tau,L} \\ \tau_L \end{pmatrix} \right), \cr
	u_R^i = \left( u_R, c_R, t_R \right), \ 
      \nu_R^i = \left( \nu_{e,R}, \nu_{\mu,R}, \nu_{\tau,R} \right), \
      d_R^i = \left( d_R, s_R, b_R \right), \ 
      \ell_R^i = \left( e_R, \mu_R, \tau_R \right).
}
\end{equation}
where $\Psi_c = C \Psi^*$ indicates charge conjugation with $C=i \sigma_2$
and 
$\epsilon_{kl} = i \sigma_2$ is the $SU(2)_L$ contraction of the two
fundamental representations, $\boldsymbol{\bar{L}}_c$ and $\boldsymbol{Q}_a$.
The explicit appearance of the charge-conjugation operator $C$ with the bilinear vertex term $\sigma_2$
gives non-vanishing vertex terms when coupling a scalar particle to a fermion pair with the same chirality.
The gauge-covariant derivatives in the kinetic energy term generate $S-\bar{S}$ pair production
diagrams for a gluon, photon or Z boson trilinear vertex.  We also expect seagull diagrams
of the S boson interacting with a pair of photons, Z bosons and gluons, respectively.
The Higgs vev will generate an S boson mass term via the $\Phi$-$S$ Yukawa interaction.

An investigation of possible $SU(5)$ representations of the $S$ boson indicates a
possible match with the charge -1$e$/3 color triplet Higgs boson
in the $\boldsymbol{5}$ or $\boldsymbol{24}$ of $SU(5)$ GUT models.  However, 
the color triplet of $SU(5)$ is known to be problematic due to the potential 
implications for proton decay.  Here, the $S$ boson Lagrangian ${\cal L}_S$ does not 
couple directly to di-quarks and appears to have only a subset of corresponding $SU(5)$ 
motivated interaction vertices.

\section{Phenomenology of ${\bf S}$ Boson Production and Decay at Hadron Colliders}

\begin{figure}
\includegraphics[width=\textwidth]{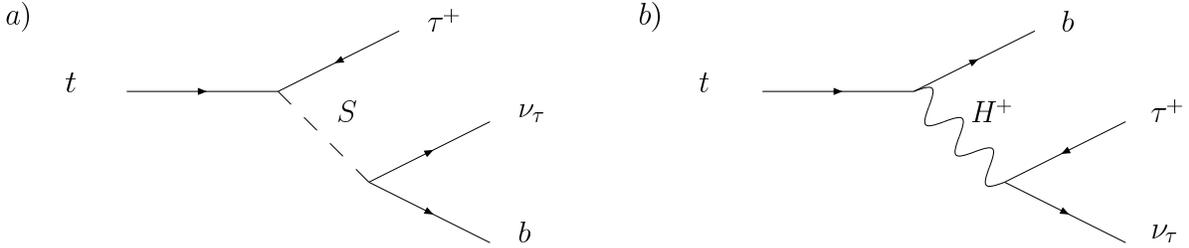}%
\caption{The decay of a top quark into a $\tau^+$ lepton and an $S$ boson is shown
on the left a).  The $S$ boson subsequently decays into a $\nu_\tau$ neutrino and
a $b$-quark.  On the right b), the search for a charged Higgs boson in top quark
decay has an identical final state with different kinematics.\label{topquarkdecay}}
\end{figure}

If we assume that the coupling of $S$ to the first two mass generations is
small in comparison to Standard Model interactions, then we can focus on the
phenomenology of the heaviest generation of quarks and leptons.  In particular,
the heaviest known particle, the top quark couples directly to the $S$ particle
and could be produced in top quark decays for values of $m_S$ lighter than the
top quark mass.  A top quark decay diagram involving the $S$ is shown 
in figure~\ref{topquarkdecay}a).  A similar final state appears in a standard
charged Higgs search, as shown in figure~\ref{topquarkdecay}b).  The main
difference in the two searches is in the kinematics of the final state.  The
charged Higgs search applied directly to a search for the $S$ boson would 
lack an explicit variable for the reconstructed $S$ mass.  This would significantly
reduce  the cross sensitivity of the charged Higgs search applied to a search 
for the $S$ boson.  However, limits on anomalous leptonic branching fractions constrained 
by the rate of $t \rightarrow \tau + X$ compared with non-$\tau$ decays would still apply.
The branching fraction searches are currently limited at the few percent level, and
would leave the possibility for $S$ production in top quark decay below a few 
percent~\cite{PDG2008}.
Similarly, top quark mass measurements using methods sensitive to the detailed
kinematics, such as the matrix method, are not normally applied to the $\tau$+four-jet
final state and would not exclude the $S$ boson directly.

\begin{figure}
\includegraphics[width=0.4\textwidth]{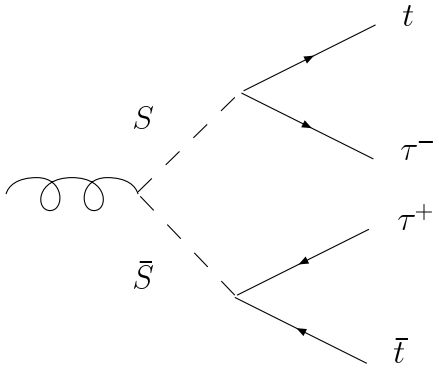}%
\caption{The $S$ boson pair production search topology
in the $t\bar{t}\tau^+\tau^-$ final state.\label{tttautau}}
\end{figure}

Another heavy-flavor search sensitive to $S$ production is pair production of 
third generation scalar leptoquarks in the channel $b\bar{b}\nu_\tau\bar{\nu}_\tau$.  
However, the search region for the third generation leptoquark has moved beyond
the kinematic threshold for $LQ3$ decays into $t \tau^-$.  If the
branching fraction is dominantly in the $t\tau$-decay mode, then the previous
searches will not be sensitive to the high-mass region~\cite{Abazov:2007bsa}.
A diagram for $S$ boson pair production
in the $t\bar{t}\tau^+\tau^-$ final state is shown in figure~\ref{tttautau}.
In the pair production search, two equal mass scalars are produced, each decaying
to a $t$-quark and a $\tau$-lepton with the corresponding charge-conjugate decay products 
for the antiparticle.

The $S$ boson pair production search in the $t\bar{t}\tau^+\tau^-$ final state 
is a new search topology for which there are no reported experimental results.
Previous searches in the $b\bar{b}\nu_\tau\bar{\nu}_\tau$ pair production topology 
were able to achieve a mass sensitivity in excess of $200$~GeV/$c^2$~\cite{Abazov:2007bsa}
with less then $1~fb^{-1}$ of integrated luminosity.
With the six-fold increase in integrated luminosity at the Tevatron and the 
well-understood status of the detectors, a search in the
$t\bar{t}\tau^+\tau^-$ final state should provide a rapid
test of the $S$ boson prediction for $m_S$ masses up to approximately $250$~GeV/$c^2$.  
For single $S$ boson production, existing top quark charged Higgs search topologies can 
be readily adapted to searches for the $S$ boson.
For higher sensitivity to rare top quark decays, the two orders of magnitude increase 
in top quark pair production at the LHC would be the highest sensitivity dataset
for $m_S$ less than the top quark once the LHC has accumulated an
integrated luminosity comparable to the Tevatron.

\section{Conclusions}

The $S$ boson prediction is an unexpected byproduct of a hypercharge splitting analysis
of the tables of known leptons and quarks.  However, there is little explanation in 
the current understanding of the Standard Model for the origin of the lepton
and quark generations and in linking the presence of a Higgs boson, needed
for spontaneous symmetry breaking, to the rest of the matter content in the
model.  The conjecture of a gauge supersymmetry in the Standard Model is a 
natural choice to generate fixed hypercharge splittings.  Given that the
$|\Delta Y|=1$ splitting is directly related to the Higgs boson, it follows
that the $|\Delta Y|=4/3$ splitting could be attributed to a new kind of scalar
particle, the $S$ boson.  Assuming the couplings of the $S$ are dominately
in the heaviest mass generation, the observation of lepton violation would be
suppressed.  Review of existing searches at hadron colliders sensitive 
to $S$ boson production highlight
similar final states present in the search for charged Higgs bosons in top quark decay.
A new search is proposed in the pair production process in the
$t\bar{t}\tau^+\tau^-$ final state.
The constraints coming from the existing searches applied to the $S$ boson would 
allow the possibility of $S$ boson production in top quark decay at the few 
percent level or less.  
The constraints on $S$ boson pair production in the high mass region from the 
$b\bar{b}\nu_\tau\bar{\nu}_\tau$ search are expected to be weak if the coupling to the $t$-quark $\tau$-lepton
final state dominates.  The existing dataset at the Tevatron and the readily available search channels
described above provide a clear path for rapid investigation of $S$ boson production.
Ultimately, rare top quark decays at the LHC will have the highest
sensitivity for the search of the $S$ boson for $m_S$ less than the top quark mass.

\section{Acknowledgements}
The author would like to thank Ben Safdi for his careful review of the Feynman rules
generated by the proposed Lagrangian.

\bibliography{newscalar}

\begin{thebibliography}{1}%
\makeatletter
\providecommand \@ifxundefined [1]{%
 \ifx #1\undefined \expandafter \@firstoftwo
 \else \expandafter \@secondoftwo
\fi
}%
\providecommand \@ifnum [1]{%
 \ifnum #1\expandafter \@firstoftwo
 \else \expandafter \@secondoftwo
\fi
}%
\providecommand \enquote [1]{``#1''}%
\providecommand \bibnamefont  [1]{#1}%
\providecommand \bibfnamefont [1]{#1}%
\providecommand \citenamefont [1]{#1}%
\providecommand\href[0]{\@sanitize\@href}%
\providecommand\@href[1]{\endgroup\@@startlink{#1}\endgroup\@@href}%
\providecommand\@@href[1]{#1\@@endlink}%
\providecommand \@sanitize [0]{\begingroup\catcode`\&12\catcode`\#12\relax}%
\@ifxundefined \pdfoutput {\@firstoftwo}{%
 \@ifnum{\z@=\pdfoutput}{\@firstoftwo}{\@secondoftwo}%
}{%
 \providecommand\@@startlink[1]{\leavevmode\special{html:<a href="#1">}}%
 \providecommand\@@endlink[0]{\special{html:</a>}}%
}{%
 \providecommand\@@startlink[1]{%
  \leavevmode
  \pdfstartlink
   attr{/Border[0 0 1 ]/H/I/C[0 1 1]}%
   user{/Subtype/Link/A<</Type/Action/S/URI/URI(#1)>>}%
  \relax
 }%
 \providecommand\@@endlink[0]{\pdfendlink}%
}%
\providecommand \url  [0]{\begingroup\@sanitize \@url }%
\providecommand \@url [1]{\endgroup\@href {#1}{\urlprefix}}%
\providecommand \urlprefix [0]{URL }%
\providecommand \Eprint[0]{\href }%
\@ifxundefined \urlstyle {%
  \providecommand \doi [1]{doi:\discretionary{}{}{}#1}%
}{%
  \providecommand \doi [0]{doi:\discretionary{}{}{}\begingroup
  \urlstyle{rm}\Url }%
}%
\providecommand \doibase [0]{http://dx.doi.org/}%
\providecommand \Doi[1]{\href{\doibase#1}}%
\providecommand \bibAnnote [3]{%
  \BibitemShut{#1}%
  \begin{quotation}\noindent
    \textsc{Key:}\ #2\\\textsc{Annotation:}\ #3%
  \end{quotation}%
}%
\providecommand \bibAnnoteFile [2]{%
  \IfFileExists{#2}{\bibAnnote {#1} {#2} {\input{#2}}}{}%
}%
\providecommand \typeout [0]{\immediate \write \m@ne }%
\providecommand \selectlanguage [0]{\@gobble}%
\providecommand \bibinfo [0]{\@secondoftwo}%
\providecommand \bibfield [0]{\@secondoftwo}%
\providecommand \translation [1]{[#1]}%
\providecommand \BibitemOpen[0]{}%
\providecommand \bibitemStop [0]{}%
\providecommand \bibitemNoStop [0]{.\EOS\space}%
\providecommand \EOS [0]{\spacefactor3000\relax}%
\providecommand \BibitemShut [1]{\csname bibitem#1\endcsname}%
\bibitem{standard_model}%
  \BibitemOpen
  \bibinfo {note} {S. L. Glashow, Nucl.\ Phys. {\bf 22} (1961) 579;\\ S.
  Weinberg, Phys.\ Rev.\ Lett. {\bf 19} (1967) 1264;\\ A. Salam, ``Elementary
  Particle Theory'', Ed. N. Svartholm, Stockholm, ``Alm\-quist and Wiksell''
  (1968), 367;\\ H. D. Politzer, Phys.\ Rev.\ Lett. {\bf 30} (1973) 1346;\\ D.
  J. Gross, F. Wilczek, Phys.\ Rev.\ Lett. {\bf 30} (1973) 1343}%
  \bibAnnoteFile{NoStop}{standard_model}%
\bibitem{higgs_mech}%
  \BibitemOpen
  \bibinfo {note} {P. W. Higgs, Phys.\ Lett. {\bf 12} (1964) 132,~Phys.\ Rev.\
  Lett. {\bf 13} (1964) 508 and Phys.\ Rev. {\bf 145} (1966) 1156;\\ F.~Englert
  and R.~Brout, Phys.\ Rev.\ Lett. {\bf 13} (1964) 321.}%
  \bibAnnoteFile{Stop}{higgs_mech}%
\bibitem{susy}%
  \BibitemOpen
  \bibinfo {note} {Y.A. Golfand and E.P. Likhtman, Sov.\ Phys.\ JETP {\bf 13}
  (1971) 323; \\ D.V. Volkhov and V.P. Akulov, Phys.\ Lett. {\bf B 46} (1973)
  109; \\ J. Wess and B. Zumino, Nucl.\ Phys. {\bf B 70} (1974) 39;\\ P. Fayet
  and S. Ferrara, Phys.\ Rep. {\bf C 32} (1977) 249;\\ A. Salam and J.
  Strathdee, Fortschr.\ Phys. {\bf 26} (1978) 57.}%
  \bibAnnoteFile{Stop}{susy}%
\bibitem{HiggsHunters}%
  \BibitemOpen
  \bibfield{author}{%
  \bibinfo {author} {\bibfnamefont{J.~F.}\ \bibnamefont{Gunion}}
  \emph{et~al.},\ }%
  \emph{\bibinfo {title} {The Higgs Hunter's Guide}}\ (\bibinfo {publisher}
  {Addison-Wesley},\ \bibinfo {year} {1990})%
  \bibAnnoteFile{NoStop}{HiggsHunters}%
\bibitem{PDG2008}%
  \BibitemOpen
  \bibfield{author}{%
  \bibinfo {author} {\bibfnamefont{C.}~\bibnamefont{Amsler}} \emph{et~al.}
  (\bibinfo {collaboration} {Particle Data Group}),\ }%
  \bibfield{journal}{%
  \Doi{10.1016/j.physletb.2008.07.018}{\bibinfo {journal} {Phys. Lett.}}\ }%
  \textbf{\bibinfo {volume} {B667}},\ \bibinfo {pages} {1} (\bibinfo {year}
  {2008})%
  \bibAnnoteFile{NoStop}{PDG2008}%
\bibitem{Abazov:2007bsa}%
  \BibitemOpen
  \bibfield{author}{%
  \bibinfo {author} {\bibfnamefont{V.~M.}\ \bibnamefont{Abazov}} \emph{et~al.}
  (\bibinfo {collaboration} {D0}),\ }%
  \bibfield{journal}{%
  \Doi{10.1103/PhysRevLett.99.061801}{\bibinfo {journal} {Phys. Rev. Lett.}}\
  }%
  \textbf{\bibinfo {volume} {99}},\ \bibinfo {pages} {061801} (\bibinfo {year}
  {2007}),\ \Eprint{http://arxiv.org/abs/0705.0812}{arXiv:0705.0812 [hep-ex]}%
  \bibAnnoteFile{NoStop}{Abazov:2007bsa}%
\end{thebibliography}%

\end{document}